
\centerline {\bf 1. Introduction}
\vskip .20in
The effects of intense magnetic fields on various phenomena, in both laboratory
and astrophysical dynamical systems, have been investigated by numerous
authors$^{[1,2]}$. Specially, a number of the principal features of
high-energy,
high-field electromagnetic conversion processes have been well studied, such as
magnetic bremsstrahlung (synchrotron radiation), magnetic pair production,
direct
and indirect trident cascades, photon splitting, and magnetic Cerenkov
radiation.
However, the effects of an intense magnetic field in astrophysical
investigations have
often been ignored. In particular, the effects on the the rates of processes
which are fundamental to studies of many astrophysical
objects (e.g., neutron stars, supernova, and etc.), and even Big Bang
Nucleosynthesis,
are usually neglected. For example, in neutron stars,
where the magnetic field may be as large as $10^{12}\sim 10^{14}\,{\rm
gauss}^{[3,4]}$, the
effects on reaction rates which are relevant to the cooling rates of neutron
star
can be substantial. A joint work of Ruderman, Cannuto, Lodenquai, and
Tsuruta on the general effects of strong magnetic fields and
superfluidity on neutron star cooling, has been presented in Ref.[5,6], where
they have specially studied the effects of magnetic fields on photon opacities.
They found that the major effect of strong magnetic fields is to
drastically reduce photon opacities in certain spacial conditions and greatly
accelerate the cooling rates. However, the influences on the cooling rates
directly from the altered URCA rates by strong magnetic fields still remain.
Moreover, in the early universe, the effects on the
weak-interaction rates which determine the rate of production of helium and
other
light elements can also be significant, depending on the possible existence of
magnetic fields at that time. In this paper, we examine the effects of a
magnetic
field on reaction
rates (including both weak and strong interactions) in the presence of variable
degrees of
degeneracy. Some potential astrophysical implications of our results will also
be
discussed.

\vskip .20in
\centerline{\bf 2. Relativistic Motion of a Charged }
\centerline{\bf Particle in a Magnetic Field}
\vskip .20in
The Dirac equation for the motion of a charged particle in an arbitrary
magnetic
field is, in conventional notation,
$${i\hbar {\partial \psi \over \partial t} = [i\hbar {\vec \balpha} \cdot
(c{\bf\vec p}
+ e{\bf \vec A}) + \gamma_4 m c^2 + e\hbar c \gamma_4 {\vec \bsigma} \cdot
{\bf\vec
B}]\psi}, \eqno(2.1)$$
where $\psi$ is the wave function of the charged particle, $e$ and $ m$ are the
electric charge and mass, ${\bf \vec A}$ is the electromagnetic vector
potential,
${\bf \vec B}$ is the strength of the magnetic
fields, $\hbar$ denotes the Planck constant, and $c$ represents the
speed of light. The Dirac operators ${\vec \balpha}$ and $\gamma_4$ are chosen
to
be of the
following forms
$${\vec \balpha} = \pmatrix{
0&{\vec \bsigma}\cr
{\vec \bsigma}&0\cr}, \quad
\gamma_4 = \pmatrix{
 I&0\cr
0& I\cr}, \quad
I = \pmatrix{
1&0\cr
0&1\cr},\eqno (2.2)$$
where ${\vec \bsigma}=(\sigma_x, \sigma_y, \sigma_z)$ is the usual Pauli spin
matrix,
with components
$$ \sigma_x = \pmatrix{
0&1\cr
1&0\cr}, \quad
\sigma_y = \pmatrix{
0&-i\cr
i&0\cr},\quad
\sigma_z = \pmatrix{
1&0\cr
0&-1\cr}.\eqno(2.3)$$
For a constant magnetic field oriented along the z axis
(${\bf\vec A}=0,\,\, {\bf\vec B} = B_z =B ={\rm const.}$), the energy
eigenvalues $E$
for the charged particle are given by
$$E^2 = c^2p^2 + m^2 c^4 + e\hbar c B\,(2n + s + 1),\eqno(2.4)$$
where $n = 0,\,1,\,2,\,\,\cdot \cdot \cdot$ is the principal quantum number,
$s$ ($=
\pm 1$ refers to spin up and spin down) is the spin variable, and $p$ is the
momentum
of the particle. The third term in the expression reflects the contribution to
the
energy of the
particle resulting from the interaction of the particle's magnetic moment with
the
magnetic field. This expression is very useful in determining the influence of
the
magnetic field on the interaction rates of many processes.
\vskip .20in
\centerline {\bf 3. Interaction Rates in the Presence of Magnetic }
\centerline {\bf Fields and Arbitrary Degeneracy}
\vskip .20in
\noindent {\bf 3.1 Weak reaction rates:}
\vskip .15in
We here consider three fundamental weak interactions which act to determine the
critical n/p ratio (see Schramm and Wagoner 1977)$^{[7]}$in many astrophysical
processes, including e.g., big bang nucleosynthesis and neutron star cooling:
$$ n \,\, + \,\, e^+ \,\, \rightleftharpoons \,\, p\,\, + \,\,\overline\nu ,
\eqno(a)$$
$$ n \,\, + \,\, \nu \,\, \rightleftharpoons \,\, p\,\, + \,\,e^- , \eqno (b)$$
$$ n \,\, \rightleftharpoons \,\, p\,\, + \,\,e^- \,\, + \,\,\overline\nu
.\eqno (c)$$
The cross-sections for these reactions can be computed using standard charged
current $\beta$ - decay theory, with the well-known V-A interaction Hamiltonian
$^{[8]}$
$$H = {g_V\over \sqrt 2} \bigr [ {\overline u_p} \gamma_\mu ( 1 - \alpha
\gamma_5)
u_n \bigr ] \bigr [ {\overline u_e} \gamma^\mu ( 1 - \gamma_5 ) {\overline
v_\nu}
\bigr ] + h.\,\, c., \eqno(3.1)$$
where $g_V = 1.4146\times 10^{-49}$erg ${\rm cm}^3$, and $\alpha = {g_A \over
g_V}
\simeq -1.262.^{[9]}$  Here ${\overline u_p},\,\,u_n,\,\,{\overline u_e},$ and
$v_\nu$
stand for the
proton, neutron, electron, and neutrino operators, with ${\overline \psi} =
\psi^{\dag}
\gamma_4$.

For the case of a constant B-field$^{[10]}$, we use the exact
relativistic wave function in a constant uniform magnetic field for an electron
and a
free-particle relativistic spinor wave function for an antineutrino. As to the
neutron
and proton, since the effects of a magnetic field on them are small compared
with
their rest mass energy difference, they can be treated nonrelativistically.
To a good approximation, we can therefore use the free-particle nonrelativistic
spinor
wave functions for them.

The energy eigenvalues for an electron are obtained from Eq.(2.4)
$$ E^2 = c^2 p^2 + m_e^2 c^4 + e \hbar c B (2n + s + 1), \eqno(3.2)$$
where $m_e$
is the rest mass of electron. In the
nonrelativistic limit, this becomes
$$ E^2 = c^2 p^2 + m_e^2 c^4 + 2 e \hbar c B n. \eqno(3.3)$$
If the electron spin is not measured, the matrix element for each of the
reactions
($(a)\to (c)$) will be the same, i.e.,
$$\mid H_{fi}(P,p,n)\mid ^2 = {g_V^2 m_e^2 c^4 \gamma \over 2\pi \hbar^2}
(1+3\alpha^2)
\{ 1-{1\over 2}\delta_{n,0} (1 - x ) + P \Lambda [x + {1\over 2}\delta_{n,0} (1
- x)]
\}, \eqno(3.4)$$
where
$$x \equiv {c p/ E},\quad \gamma \equiv {1\over 2}{B/ B_c},\quad \Lambda
\equiv 2\alpha (1 - \alpha) / {(1 + 3\alpha^2)} \simeq - 0.99,\eqno(3.5)$$
and the subscripts $f\,\,{\rm and}\,\,i$ denote, respectively, the final and
initial
states. $P$ is the polarization of the
neutron source( $P=0$: unpolarized; $P=1$: completely polarized; $0< P < 1$:
partially
polarized). $B_c = {m_e^2 c^3\over e\hbar} = 4.4 \times 10^{13} {\rm gauss}\,$
is the
field strength at which magnetic quantum levels, ``cyclotron lines," begin to
occur$^{[11]}$.

The reaction rate is obtained from Fermi's {\it golden rule} (cf 12)
$$\lambda_{i \to f} = {2\pi \over \hbar} \sum_i n_i \int_f dE_f {dn_f\over
dE_f}\delta(E_f - E_i) \mid H_{fi}(P,p,n)\mid ^2, \eqno(3.6)$$
where $n_i$ is the distribution of initial states and ${dn_f\over dE_f}$ is the
density of final states.

For convenience, we introduce the following parameters:
$$\epsilon \equiv {E\over m_e c^2}, \quad q\equiv {m_n - m_p \over m_e}, \quad
Z_e
\equiv {m_e c^2\over kT_e},$$
$$Z_\nu \equiv {m_e c^2\over kT_\nu}, \quad \phi_e \equiv {\mu_e \over kT_e},
\quad
\phi_\nu \equiv {\mu_\nu \over kT_\nu}, \eqno(3.7)$$
where $m_n$ and $m_p$ are the rest masses of the neutron and proton,
respectively,
$T_{e\, {\rm or}\, \nu}$ represents the temperature of the electron or
neutrino,
$\mu_{e\,{\rm or}\,\nu}$ is the chemical potential of the electron or neutrino,
and
$\phi_i (i=e,\,\nu)$ is the degeneracy parameter.

Thus, the rate for reaction $n + e^+ \to p + \overline\nu_e$, is given by
$$\eqalign{\lambda_a =& {g_V^2 (1 + 3\alpha^2) m_e^5 c^4 \gamma \over 2 \pi^3
\hbar^7}
\sum_{n=0}^{\infty} [2-\delta_{n0}(1-P\Lambda)] \int_{\sqrt {1+4\gamma
n}}^\infty
{\epsilon d\epsilon \over \sqrt{\epsilon^2-(1+4\gamma n)}}\cr
&\qquad \qquad \times {1\over (1+e^{\epsilon Z_e + \phi_e})} {(\epsilon + q)^2
e^{(\epsilon + q) Z_\nu + \phi_\nu} \over (1+e^{(\epsilon + q) Z_\nu +
\phi_\nu})}.\cr}\eqno(3.8)$$
Similarly, for reaction (b), $n + \nu \to p + e^-$, the rate is
$$\eqalign{\lambda_b =& {g_V^2 (1 + 3 \alpha^2) m_e^5 c^4 \gamma \over 2 \pi^3
\hbar^7}
\sum_{n=0}^{\infty} [2-\delta_{n0}(1-P\Lambda)] \int_{\sqrt {1+4\gamma
n}}^\infty
{\epsilon d\epsilon \over \sqrt{\epsilon^2-(1+4\gamma n)}}\cr
&\qquad \qquad \times {1\over (1+e^{\epsilon Z_e + \phi_e})} {(\epsilon - q)^2
e^{\epsilon Z_e + \phi_e} \over (1+e^{(\epsilon - q) Z_\nu - \phi_\nu})}\cr
&- {g_V^2 (1 + 3 \alpha^2) m_e^5 c^4 \gamma \over 2 \pi^3 \hbar^7}
\sum_{n=0}^{n_{\rm
max}} [2-\delta_{n0}(1-P\Lambda)] \int_{\sqrt{1+4\gamma n}}^q {\epsilon
d\epsilon \over
\sqrt{\epsilon^2-(1+4\gamma n)}}\cr
&\qquad \qquad \times {1\over (1+e^{\epsilon Z_e + \phi_e})} {(\epsilon - q)^2
e^{\epsilon Z_e + \phi_e} \over (1+e^{(\epsilon - q) Z_\nu - \phi_\nu})}.\cr}
\eqno(3.9)$$
where $n_{\rm max}$ is the largest integer in $(q^2 -1)/4\gamma$ and $q =
2.53$.
Finally, for reaction (c), $n \to p + e^- + \overline\nu_e$, we have
$$\eqalign{\lambda_c =& {g_V^2 (1 + 3 \alpha^2) m_e^5 c^4 \gamma \over 2 \pi^3
\hbar^7}
\sum_{n=0}^{n_{\rm max}} [2-\delta_{n0}(1-P\Lambda)] \int_{\sqrt{1+4\gamma
n}}^q
{\epsilon d\epsilon \over \sqrt{\epsilon^2-(1+4\gamma n)}}\cr
&\qquad \qquad \times {1\over (1+e^{\epsilon Z_e + \phi_e})} {(q-\epsilon)^2
e^{\epsilon Z_e + \phi_e } e^{(q-\epsilon) Z_\nu + \phi_\nu} \over
(1+e^{(q-\epsilon)
Z_\nu + \phi_\nu})}.\cr} \eqno(3.10)$$
{}From these expressions, we can clearly see that, if
$\gamma \ge (q^2 -1)/4 \simeq 1.35,$ (or, equivalently,  $B>10^{14}$gauss),
${\rm then}\,\, n_{\rm max} = \big[{q^2-1\over 4\gamma}\big] = 0$, and the rate
of
$\beta$-decay would behave linearly with
$B$. In addition, the rate $\lambda_c$ (Eq.(3.10)) is exactly cancelled by the
second
term of $\lambda_b$ (Eq.(3.9)) if $\phi_\nu$ is negligible.

Finally, we are interested in the total reaction rates for converting neutrons
into
protons (and vice-versa), which can be obtained by summing the above three
rates
$$\eqalign{\lambda_{n\,\to\,p} &= \lambda_a \quad + \quad \lambda_b \quad +
\quad
\lambda_c \cr
&= {g_V^2 (1 + 3 \alpha^2) m_e^5 c^4 \gamma \over 2 \pi^3 \hbar^7}
\sum_{n=0}^{\infty}
[2-\delta_{n0}(1-P\Lambda)] \cr
& \times \int_{\sqrt {1+4\gamma n}}^\infty \epsilon d\epsilon
{[\epsilon^2-(1+4\gamma n)]^{-{1\over 2}}\over (1+e^{\epsilon Z_e + \phi_e})}
\bigr
[{(\epsilon+q)^2 e^{(\epsilon+q) Z_\nu  + \phi_\nu} \over 1+e^{(q+\epsilon)
Z_\nu +
\phi_\nu}} + {(\epsilon-q)^2 e^{\epsilon Z_e + \phi_e}\over
1+e^{(\epsilon-q)Z_\nu -
\phi_\nu}}\bigr ].\cr} \eqno(3.11)$$
For illustrative purpose, we will briefly examine several limiting cases.

\noindent{\it 1. No magnetic fields:}

In the absence of a magnetic field, $B=0,$ which means $\gamma = 0$ and $P =
0$. Thus,
the summation over the principal
quantum number $n$ needs to be replaced by an integration. We define a variable
$$\theta \equiv 1 + 4 \gamma n, \eqno(3.12)$$
such that
$$\sum_n^\infty \to \int_0^{\infty} dn = \lim_{\gamma\to0}\,\,{1\over
4\gamma}\int_1^\infty d\theta.\eqno(3.13)$$
Substituting Eq.(3.13) into Eqs.(3.8), (3.9), (3.10), and (3.11), and
integrating
over the variable $\theta$, we finally obtain
$$\lambda_a ( B=0 ) =\quad {1\over \tau} \int_1^\infty {\epsilon d\epsilon
\sqrt{\epsilon^2 -1} \over (1+e^{\epsilon Z_e + \phi_e})} {(\epsilon + q)^2
e^{(\epsilon + q )Z_\nu + \phi_\nu}\over
(1+e^{(q+\epsilon) Z_\nu + \phi_\nu})}, \eqno(3.14)$$
$$\eqalign{\lambda_b ( B=0 ) =& \quad {1\over \tau} \int_1^\infty {\epsilon
d\epsilon \sqrt{\epsilon^2-1} \over (1+e^{\epsilon Z_e + \phi_e})} {(\epsilon -
q)^2
e^{\epsilon Z_e + \phi_e } \over (1+e^{(\epsilon - q ) Z_\nu - \phi_\nu})}\cr
&- {1\over \tau} \int_1^q {\epsilon d\epsilon \sqrt{\epsilon^2-1} \over
(1+e^{\epsilon Z_e + \phi_e})} {(\epsilon -q)^2 e^{\epsilon Z_e + \phi_e} \over
(1+e^{(\epsilon - q) Z_\nu - \phi_\nu})}.\cr} \eqno(3.15)$$
$$\lambda_c ( B=0 ) = \quad {1\over \tau} \int_1^q {\epsilon d\epsilon
\sqrt{\epsilon^2-1} \over (1+e^{\epsilon Z_e + \phi_e})} {(q-\epsilon)^2
e^{\epsilon
Z_e + \phi_e } e^{(q-\epsilon) Z_\nu + \phi_\nu}\over (1+e^{(q-\epsilon) Z_\nu
+
\phi_\nu})}, \eqno(3.16)$$
and
$$\eqalign{\lambda_{n\to p}(B=0) =& {1\over \tau} \int_1^\infty {\epsilon
d\epsilon
\sqrt{\epsilon^2 - 1} \over (1+e^{\epsilon Z_e + \phi_e})} \bigr [
{(q+\epsilon)^2
e^{(\epsilon + q )Z_\nu + \phi_\nu}\over (1+e^{(q+\epsilon) Z_\nu +
\phi_\nu})}\cr
&\qquad \qquad \qquad \quad + {(\epsilon-q)^2 e^{\epsilon Z_e + \phi_e} \over
(1+e^{(\epsilon-q)Z_\nu-\phi_\nu})}\bigr ],\cr} \eqno(3.17)$$
where ${1\over \tau} \equiv {g_V^2 (1 + 3 \alpha^2) m_e^5 c^4 \over 2 \pi^3
\hbar^7} \simeq 6.515 \times 10^{-4} {\rm sec}^{-1}.$ We express our results
relative
to the measured life-time $\tau_n$ of the neutron ($\tau_n \sim 889.6 \pm 2.9
\, {\rm
sec.},$ $ \tau_{1/2} \sim 10.277 \pm 0.046 \,{\rm min.})^{[13]}$,
$$\tau = I\,\tau_n,$$
where $I$ is the value of the integral.

These formulae are merely the generalized form of the rates derived by Peebles
and Wagoner et al$^{[9]}$ and used in standard Big Bang Nucleosynthesis (BBN)
calculations.$^{[9]}$ Thus our formalism reduces to the correct limit as $B \to
0$.

\noindent {\it 2. Weak magnetic fields} ($ B\ll B_c, \, \,{\rm that\,\, is}\,\,
\gamma
\ll 1) :$

In this case, we can approximate
$$( \epsilon^2 - 1 - 4 \gamma n )^{-1/2} \simeq ( \epsilon^2 - 1 )^{-1/2} + 2
\gamma n
( \epsilon^2 - 1 )^{-3/2} + 0(\gamma^2). \eqno(3.18)$$
Incorporating this, together with equation (3.13), into equations (3.8), (3.9),
(3.10), and (3.11), and replacing the
summation over $n$ by an integral, we obtain the result
$$\lambda_i (B\ll B_c) \simeq \lambda_i (B=0) \bigr [ 1 + 0(\gamma^2)\bigr ].
\eqno(3.19)$$
where $i = a, \,b, \,c, \, {\rm of}\,\, n\to p$. Thus, it can be seen that a
weak
magnetic field $( B\ll
B_c)$ will have a negligible effect on the rates of the weak reactions.

\noindent {\it 3. Strong magnetic fields} ($ B\gg B_c, \, \, \gamma \gg 1 ):$

If the magnetic field is strong, $\gamma \gg 1$, then $n_{\rm max} \to 0$,
which
implies that there is only one term $(n=0)$ left in the summation over $n$. In
order to
see the effect of the magnetic field, we rewrite the first function in the
integrals
as
$$( \epsilon^2 - 1 - 4 \gamma n )^{-1/2} \simeq { ( \epsilon^2 - 1 )^{1/2}
\over \epsilon^2 - 1}. \eqno(3.20) $$
Since $\epsilon$ goes from 1 to $\infty$, in a mathematical sense, it is always
true
that
$$ \int_1^\infty d\epsilon \,{ ( \epsilon^2 - 1 )^{1/2} \over \epsilon^2 - 1}\,
g(\epsilon)\biggr|_{B\not= 0} \le 2 \int_1^\infty d\epsilon \, ( \epsilon^2 - 1
)^{1/2}\, g(\epsilon)\biggr |_{B=0}. \eqno(3.21)$$
where $g(\epsilon)\, ( \ge 0,\,\, {\rm for\,\, any }\,\,\epsilon )$ is an
arbitrary positive exponential decay function. Thus, by comparing with the
rates
for zero magnetic field, we have
$$\lambda_i (B\gg B_c) \le 2 \gamma\, \lambda_i (B=0)\,\, ( > \lambda_i (B=0)).
\eqno(3.22)$$
The dependence of the ratio of
the total neutron-depletion rate to the free-field rate on the magnetic field
parameter and the temperature, in the case of nondegeneracy
and nonpolarization, is plotted in Figure 1.
%
{}From this, we see that the effects of a strong magnetic field are negligible
until
the temperature drops to the point at which neutron $\beta$-decay begins to
dominate
and large deviations can occur.
We therefore conclude that a strong magnetic field can, in some domains, have a
significant effect on the rates of the weak reactions.

Here, it is worthwhile to point out that, if the magnetic field is not uniform,
but
rather is characterized by
a distribution function with regions(or magnetic bubbles) (${\bf \vec A}\not=0$
in Eq.(2.1)), the nuclear
reaction rates will become inhomogeneous; i.e., the field will vary with
spacial
variables, and the reaction rates will be different, region by region. Thus a
fluctuation in the reaction rate would occur (similar to the introduction of
the
density fluctuations associated with a first order QCD phase
transition$^{[14]}$).

\vskip .20in
\noindent {\bf 3.2 Strong reaction rates}
\vskip .20in
Let us now discuss the effects of magnetic fields on strong interactions. In
this
situation, the nucleons(p and n) and pions($\pi^+, \pi^0, \pi^-$) will
interact both with themselves, and with the magnetic fields, through their
charges and magnetic moments. Among these interactions, the relative
interaction
strength between themselves ($\alpha_s\sim 1$)
is about two orders of magnitude larger than that of the electromagnetic
interaction ($\alpha_e\sim
10^{-2}$) and five orders of magnitudes larger than that of the weak
interaction($\alpha_w \sim
10^{-5}$). The entirety of the phenomena caused by the strong interaction,
for example, the processes below
$$p\,\, +\,\,p \,\,\rightarrow \,\, D\,\,+\,\,\pi^+,\eqno(3.23a)$$
$$n\,\, +\,\,p \,\,\rightarrow \,\, D\,\,+\,\,\gamma,\eqno(3.23b)$$
$$n\,\, +\,\,\pi^+ \,\,\rightarrow \,\, p,\eqno(3.23c)$$
$$n\,\, +\,\,p \,\,\rightarrow \,\, D\,\,+\,\,\pi^0,\eqno(3.23d)$$
$$3\,\alpha\,\, \rightarrow \,\, ^{12}C,\eqno(3.23e)$$
is described by the quantum chromodynamic(QCD) gauge theory. For these
systems in the presence of a magnetic field, the Hamiltonian of the particles
is
composed of two parts:
$$H\, = H_0\, + \, H_I$$
where
$$H_0 = cp + m c^2\eqno(3.24)$$
and
$$H_I \sim {1\over mc} e{\bf \vec A}\cdot {\bf \vec p} + {e\hbar \over 2}
{\vec\bmu}\cdot {\bf \vec B}, \eqno(3.25)$$
respectively, represent the energy of the particle in the field-free case and
the
coupling
energy of the field to the magnetic moment ${\vec \bmu}$ of the particle. If
the
magnetic field is uniform, ${\bf\vec A}=0$, the interaction Hamiltonian will
become
$$H_I \sim  {e\hbar \over 2}{\vec\bmu}\cdot {\bf \vec B}. \eqno(3.26)$$
Since all baryons and pions involving the strong interactions have mass $m \ge
10\,m_e$ and magnetic moment $\mid{\vec \bmu}\mid \le \mid{\vec \bmu_p}\mid\sim
{1\over 660} \mid{\vec \bmu_e }\mid$, where $\mid{\vec \bmu_p }\mid$
and $\mid{\vec \bmu_e }\mid$ stand for the magnetic moment of proton and
electron
respectively, the
interaction energy between the magnetic field and the magnetic moment of the
baryons and
pions will be
$$H_I \le {\vec \bmu_p}\cdot {\bf \vec B} \sim 10^{-3}{\vec \bmu_e}\cdot {\bf
\vec
B} \sim 10^{-6} {B\over B_c}m_pc^2,\eqno(3.27)$$
where, again $B_c = 4.4 \times 10^{13}$gauss. It is apparent that this
interaction
energy is much much smaller than the rest mass energy difference of the
associated reactions, if the magnetic field is weaker than $10^{18}$ gauss.

We thus conclude that, for all magnetic fields of interest ($B\le 10^{18}$
gauss), the
contribution of the interaction energy between nucleons and magnetic fields to
the
total energy of the system is too small to be significant, and therefore that
the
effects
of magnetic fields ($B\le 10^{18}$G) on the strong interaction rates are
negligible.
\vskip .20in
\noindent {\bf 3.3 URCA and modified URCA rates}
\vskip .20in
At very high temperatures ($T\ge 10^9$K), e.g., in the core of some massive
stars, the so-called URCA rates$^{[15]}$
$$ n\,\,\, \rightarrow \,\, p\,\, +\,\, e^-\,\, +\,\, \overline\nu_e,\,\,
\eqno(3.28a)$$
$$ e^-\,\,+\,\,p\,\, \rightarrow \,\, n\,\, +\,\,\nu_e,\,\, \eqno(3.28b)$$
and modified URCA rates$^{[13]}$
$$n\,\, + \,\, n \,\,\rightarrow
\,\,n\,\,+\,\,p\,\,+\,\,e^-\,\,+\,\,\overline\nu_e,\eqno(3.29a)$$
$$ n\,\,+\,\,p\,\,+\,\,e^-\,\, \rightarrow \,\, n\,\, +\,\,n\,\,+\,\,\nu_e,\,\,
\eqno(3.29b)$$
$$n\,\, + \,\, \pi^- \,\,\rightarrow
\,\,n\,\,+\,\,e^-\,\,+\,\,\overline\nu_e,\eqno(3.29c)$$
$$n\,\, + \,\, n \,\,\rightarrow
\,\,n\,\,+\,\,p\,\,+\,\,\mu^-\,\,+\,\,\overline\nu_\mu,\eqno(3.29d)$$
$$ n\,\,+\,\,+\,\,p\,\,+\,\,\mu^-\,\, \rightarrow \,\, n\,\,
+\,\,n\,\,+\,\,\nu_\mu,\,\,\eqno(3.29e)$$
(as well as the inverse processes) provide a dominant mode of energy loss via
neutrino emission. A typical example of this will be the essential cooling
processes
involved in the interiors of neutron stars.

Ruderman et al$^{[5,6]}$ have explored the effects of the strong magnetic field
on the photon opacities, and in turn, on the cooling rates. Using our
calculations
in Section 3.1 of the weak interaction rates in the presence
of the magnetic field, we are now in a position to examine the rate of the URCA
reactions
and the modified URCA reactions in the presence of a intense magnetic field.
This
should lead to important consequences to the cooling of neutron stars since it
has
been inferred that a strong magnetic field ($\ge 10^{13}$gauss) may exist in
the
interior of the neutron stars. A detailed calculation will be presented in a
coming paper.
\vskip .20in
\centerline {\bf 4. Astrophysical Applications}
\vskip .20in
In section 3, we have derived the interaction rates as a function of magnetic
field B
in the presence of variable degrees of degeneracy and polarization. Our
calculations
have
shown that the effects of the magnetic fields on weak reaction rates are
significant,
if the strength of the magnetic field is comparable with the critical B field
$B_c = 4.4\times 10^{13}$gauss where quantized cyclotron states begin to exist.
These results can lead to important
astrophysical applications.
\vskip .20in
{\it (a) Big Bang Nucleosynthesis}

If an intense primordial magnetic field existed in the early universe,
particularly at
or just before the epoch of primordial nucleosynthesis($\sim 1 {\rm min}. $),
then the
direct influence
on the nuclear reaction rates can be consequential and, in turn, the abundances
of
light elements produced in BBN can be significantly affected. This phenomenon
will
enable us
to constrain the strength of the primordial magnetic field more accurately by
using
big bang nucleosynthesis. Detailed discussion and numerical calculations
relevant to
this issue will be presented in a separate paper$^{[16]}$.
\vskip .20in
{\it (b) Physics of astrophysical compact objects}

If compact objects (e.g., neutron stars, pulsars, and white dwarfs) indeed
have intense magnetic fields, as has been implied by a number of authors and
observed
for a number of objects, then the
effects of the fields on the fundamental physical processes involved in these
objects, which have often been neglected in earlier studies (e.g., of the
(URCA)
cooling processes of the neutron stars), should be fully taken into account.
The results presented in this paper
reveal that, if the magnetic fields in neutron stars
are strong enough (as large as $10^{12}-10^{14}$gauss), the effects of the
fields on
URCA rates may be substantial and must surely be
considered in calculations of the
cooling problem of neutron stars. A more detailed
examination of this question will appear as a subsequent paper.
\vskip .20in
\centerline {\bf Acknowledgements}
We would like to thank A. V. Olinto and D. M. Mihalas for very
helpful discussions. This research was supported in part by the NSF grants
AST-93-96039 and AST-92-17969 at the University of Chicago, in part by
NSF grant AST-90-22629, DOE grant
DE FG0291ER40606 and NASA grant NAGW 1321 at the University
of Chicago, and in part by the DOE and by NASA through grant NAGW 2381 at
Fermilab.

\vskip .20in
\centerline{\bf REFERENCES}
\vskip .20in
\par \ref
[1] J. Erber, Rev. Mod. Phys., 38 (1966) 626.
\par \ref
[2] J. J. Eklein, Rev. Mod. Phys., 40 (1968) 523.
\par \ref
[3] C. Woltjer, Astrophys. J., 140 (1964) 1309; J. Trumper, et al, Astrophys.
J.
Lett., 219 (1978) L105.
\par \ref
[4] W. A. Wheaton et al, Nature, 282 (1979) 240.
\par \ref
[5] S. Tsuruta, V. Cannuto, J. Lodenquai, and Ruderman, Astrophys. J., 176
(1972) 739.
\par \ref
[6] S. Tsuruta, in  Physics of Dense Matter, Edited by C. J. Hansen (D. Reidel
Publishing Company, Boston, U.S.A., 1974), p209.
\par \ref
[7] D. N. Schramm and R. V. Wagoner, Ann. Rev. Nucl. \& Part. Sci., 1977.
\par \ref
[8] C. S. Wu and S. A. Moszkowski, Beta Decay, (Wiley--Inter--Science, Inc.,
New
York, 1966); G. K\"all\'en, Elementary Particle Physics (Reading, Mass.:
Addison -
Wesley Publishing Ca. 1964); J. D. Bjorken and S. D. Drell, Relavistic Quantum
Mechanics (New York: McGraw-Hill Book Co., 1964).
\par \ref
[9] P. J. E. Peebles, Astrophys. J., 146 (1966) 542; R. V. Wagoner, W. A.
Fowler, and
F. Hoyle, Astrophys. J., 148 (1966) 3; R. A. Alpher, J. W. Follin, and R. C.
Herman,
Phys. Rev., 92 (1953) 1347.
\par \ref
[10] M. H. Johnson and B. A. Lippmann, Phys. Rev. 76 (1949) 828; 77 (1950) 702;
R. F.
O'Connell, Phys. Rev., 176 (1968) 1433.
\par \ref
[11] L. D. Landau and E. M. Lifshitz, The Classical Theory of Fields (Addison -
Wesley
Publishing Company, Inc., Reading Mass., 1965), revised 2nd ed., Chap. 9.
\par \ref
[12] G. Baym, Lectures on Quantum Mechanics (Benjamin/Cummings Publishing
Company,
Inc., Menlo Park, Ca. 1973).
\par \ref
[13] T. P. Walker, G. Steigman, D. N. Schramm, K. A. Olive, and H-S. Kang,
Astrophys
J., 373 51 (1991).
\par \ref
[14] H. Hurki-Suonio, R. A. Matzner, K. A. Olive, and D. N. Schramm, Astrophys.
J.,
353 (1990) 406; J. H. Applegate, C. J. Hogan, and R. J. Scherrer, Phys. Rev. D,
35
(1987) 1151.
\par \ref
[15] S. L. Shapiro and S. A. Teukolsky, Black Holes, White Dwarfs, and Neutron
Stars,
(A Wiley - Interscience Publication, John Wiley \& Sons, Inc., New York, 1983)
p. .
\par \ref
[16] B. Cheng, D. N. Schramm, and J. W. Truran, in press
\vfill
\eject
\centerline{\bf Figure Captions}
\vskip .28in
\noindent Fig. 1 The ratio of the neutron depletion rate with $B\not=0$ to
$B=0$.

\end